\documentclass[useAMS,usenatbib]{mn2e}
\usepackage{graphics}
\usepackage{graphicx}
\usepackage{bm}
\usepackage{float}
\usepackage{epsf}
\usepackage{color}

\usepackage{graphics}
\usepackage{float, tikz, subfigure, array, caption}
\usepackage{pgf}
\usepackage{mathrsfs}
\usepackage{amssymb,amsmath}
\usepackage{tabularx}

\newcommand{\myquote}[1]{``#1"}

\newcommand{\msun}{{h$^{-1}$M$_{\odot}$}}
\newcommand{\mpc}{{h$^{-1}$Mpc}}

\title[A Machine Learning Approach to Galaxy-LSS Classification I]{A Machine Learning Approach to Galaxy-LSS Classification I: Imprints on Halo Merger Trees}

\author[Hui et al.]{Jianan Hui$^{1}$\thanks{E-mail:jhui003@ucr.edu}, Miguel A. Aragon-Calvo$^{2}$, Xinping Cui$^{1}$, James M. Flegal$^{1}$ \\
$^{1}$Department of Statistics, University of California, Riverside, Riverside, CA, United States. \\
$^{2}$Instituto de Astronomia, Universidad Nacional Autonoma de Mexico, Apartado postal 877, C.P. 22800, Ensenada, B.C. Mexico\\}
\begin{document}

\date{Submitted to MNRAS}

\pagerange{\pageref{firstpage}--\pageref{lastpage}} \pubyear{2017}
\maketitle
\label{firstpage}

\begin{abstract}

The cosmic web plays a major role in the formation and evolution of galaxies and defines, to a large extent, their properties. However, the relation between galaxies and environment is still not well understood. Here we present a machine learning approach to study imprints of environmental effects on the mass assembly of haloes. We present a galaxy-LSS machine learning classifier based on galaxy properties sensitive to the environment. We then use the classifier to assess the relevance of each property. Correlations between galaxy properties and their cosmic environment can be used to predict galaxy membership to void/wall or filament/cluster with an accuracy of $93\%$. Our study unveils environmental information encoded in properties of haloes not normally considered directly dependent on the cosmic environment such as merger history and complexity. Understanding the physical mechanism by which the cosmic web is imprinted in a halo can lead to significant improvements in galaxy formation models. This is accomplished by extracting features from galaxy properties and merger trees, computing feature scores for each feature and then applying support vector machine to different feature sets. To this end, we have discovered that the shape and depth of the merger tree, formation time and density of the galaxy are strongly associated with the cosmic environment. We describe a significant improvement in the original classification algorithm by performing LU decomposition of the distance matrix computed by the feature vectors and then using the output of the decomposition as input vectors for support vector machine.

\end{abstract}

\begin{figure}
\centering
\includegraphics[width=0.45\textwidth,angle=0.0]{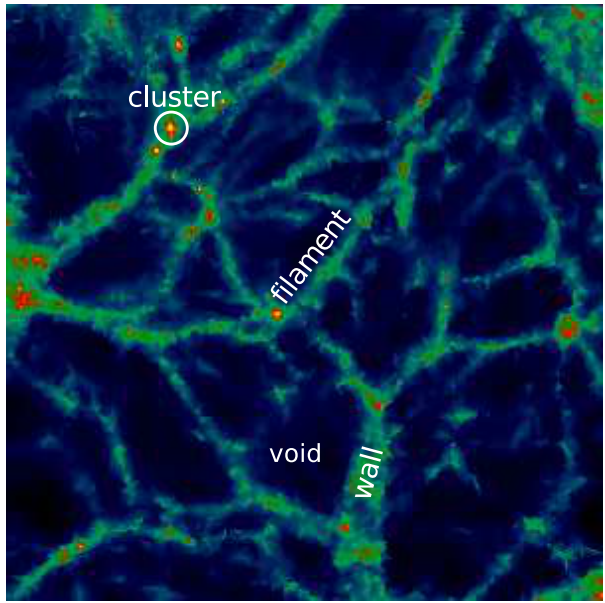}
\caption{Density field computed from an N-body simulation with a box size of 32 h$^{-1}$Mpc. The slice is 1 h$^{-1}$Mpc thick across the $z$ axis. We highlight the location of a representative cluster, filament, wall and void. Walls are two-dimensional sheets and here appear as projections. Note how voids are surrounded by wall/filaments and clusters are connected by filaments.}
\label{galaxy}
\end{figure}

%
\section{Introduction}

Understanding the relation between galaxies and their environment is a key component in a complete model of galaxy formation/evolution. This requires the ability to characterize galaxies and their cosmic environment in both a qualitative and quantitative way.  While an extensive literature exists on both galaxies and the large-scale structure (LSS)  just recently studies began to focus on their interrelation and the imprints left by the cosmic environment on galaxy properties.  In order to understand the galaxy-LSS relation, we must first be able to characterize both galaxies and their environment. 

In the last two decades, several techniques have been developed to characterize the LSS based on local variations of the matter distribution \citep{AragonMMF, Forero-Romero09, Choi10, Gonzalez10}, point processes \citep{Stoica05} and the topology of the density field \citep{Novikov06,  Platen07, Sousbie08, Aragon10a, Sousbie11}. These techniques work from either a discrete point distribution (representing mass particles galaxies or haloes) or a continuous density field derived from it. In particular, approaches based on haloes/galaxies ignore their dependence on properties with their local cosmic environment, acting only as sampling points of the underlying density field. This intrinsic assumption can potentially leave out important information that environment may have imprinted on halo/galaxies properties. Also, most standard LSS analysis techniques require significant computational resources making the analysis of large-scale N-body simulations (now of the order of trillions of particles inside large computational boxes \citep{Skillman14, Potter17}) a computational challenge.

\subsection{The galaxy-LSS connection}
Galaxies in the Universe display a wide range of properties, from blue star-forming spiral and irregular galaxies to red and quenched ellipticals. The origin of this variation is the result of  complex processes affecting galaxy evolution such as galaxy-galaxy encounters and mergers \citep{mulchaey1999isolated}, mass accretion via cold flows \citep{dekel2006galaxy,dekel2009cold}, tidal disruption \citep{byrd1990tidal}, etc. Several observations point to the cosmic environment as a key factor in galaxy evolution by defining local matter geometry and dynamics. The most salient example of the effect of environment in galaxies is perhaps the morphology/color-density relation \citep{dressler1980galaxy} describing the change in morphology/color of galaxies as a function of local density from blue spirals in low-density environments to red ellipticals in dense regions. 

The cosmic environment is the result of the anisotropic collapse of primordial matter fluctuations.  As described in \citet{Zeldovich70},  the gravitational collapse of a cloud of matter follows a succession of dynamical and geometrical stages during which its density increases as its dimensionality is reduced. This gives rise to the foam-like system known as the \myquote{cosmic web}, composed of spherically symmetric clusters, joined by thin elongated filaments which form two-dimensional membranes or walls. Clusters, filaments, and walls define the boundaries of vast empty regions or voids. Each cosmic environment has specific geometries and dynamics, with characteristic densities increasing in the order voids-walls-filaments-clusters (see Fig. \ref{galaxy}). 

\begin{figure}
\centering
\includegraphics[width=0.49\textwidth,angle=0.0]{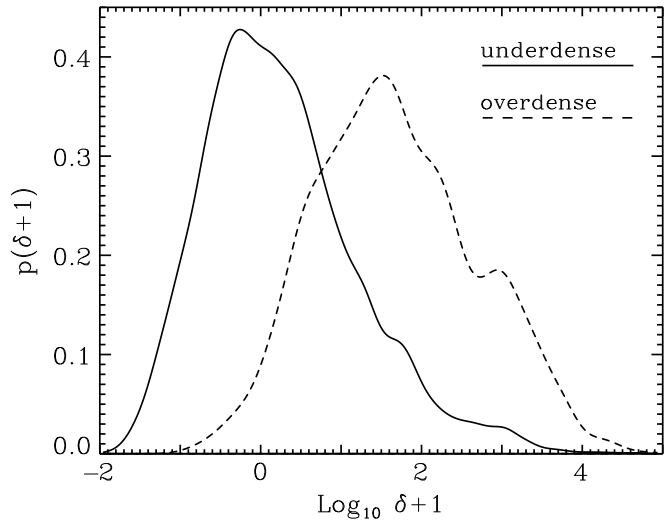}
\caption{Probability density distribution of density (Log$_{10}(\delta+1)$ for haloes in underdense regions (voids/walls, solid line) and overdense regions (filaments/clusters, dashed line). Both distributions were normalized to unitary area.}
\label{density}
\end{figure}

Voids and walls are dynamically young environments with zero and one full gravitational collapse respectively. They are characterized by low densities ($\delta < 1$) and a low-rate of galaxy-galaxy interactions. In contrast filaments and clusters (with two and three gravitational collapses respectively), having high densities ($\delta > 10$) and complex dynamics, are dominated by non-linear interactions  (see Fig. \ref{density}).

%
\subsection{Searching for imprints of environment on halo properties}

The observed galaxy distribution defining voids, walls, filaments, and clusters is determined by density fluctuations of the order of several $(\sim 4-8)$ Mpc  \citep{Einasto11}. On the other hand, galaxies originate from the collapse of a Lagrangian volume with an equivalent spherical radius of the order of $\sim 0.1-1$ Mpc. The gravitational collapse of overdensities from which galaxies emerge is modulated by the large-scale fluctuations producing the Cosmic Web. Galaxies sitting on top of large peaks will collapse earlier and have more interactions with other galaxies than galaxies in underdense regions. We should expect to find imprints of the different environments where galaxies formed in their properties \citep{Gao05, Hahn07, Hahn09, Fakhouri09, Paranjape17}. 
While there is no univocal relation between galaxies and their environment (i.e. galaxies in the same environment can have very different properties \citep{Peebles01}), we should find general trends. 
Identifying such trends and their physical origin is a challenging problem since we do not fully understand how haloes are related to their environment. One possibility is to apply techniques that do not require a full understanding of the underlying variables of the system but can still predict its behavior. In the following sections, we will describe a class of analysis techniques from the area of Machine Learning (ML) which can be used to predict complex systems and even help to understand the interplay between the variables in the system.

Observations and computer simulations point to a clear dependence of halo/galaxy properties with their cosmic environment, yet to date, all available LSS classification algorithms leave out this important information. Here we explore a new approach to the LSS classification problem by using the properties of haloes to characterize their cosmic environment, in effect turning around the standard approach of studying the properties of haloes/galaxies as a function of their cosmic environment. This exercise not only can provide us with an alternative method for LSS classification but also show give us insight on how the LSS affects halo/galaxy evolution. 

%
\subsection{Cosmic environment classes}

In this paper, we consider two classes of cosmic environments based on their dynamical state and characteristic densities: voids/walls (underdense class) being dynamically young and filaments/clusters (overdense class) being dynamically more evolved.  The division, while ignoring particular differences between the four basic cosmic environments encodes the observed relation between haloes and their environment in a similar way as the commonly used \textit{cluster vs. field} classification used to separate haloes in clusters from the rest.

%
\subsection{Machine Learning in astronomy}

There has been a significant increase in recent years in the number of studies applying ML techniques in astronomy. This has been motivated by new computational methods, faster hardware, and availability of large datasets. One of the most important applications of ML in astronomy is the determination of redshifts of galaxies from a set of broad band filters. Using ML techniques it is possible to obtain \textit{photometric redshifts} for a large number of galaxies using a few broad band filters instead of the more expensive (and accurate) spectroscopic redshifts \citep{Benitez00,Carliles10,Menard13,Cavuoti15,Sadeh15,Hoyle15,Hoyle16}. Other uses of ML in astronomy include the study of the structure of the Milky Way \citep{Riccio15} and its mass \citep{McLeod16} as well as the masses of larger systems \citep{Ntampaka15,Ntampaka15b}, the assignment of galaxies to dark matter haloes to generate mock catalogs from N-body simulations \citep{Xu13,Kamdar16,Kamdar16b} and galaxy morphological classification, a task where humans used to be the best classifiers \citep{Huertas11, Kuminski14, Schutter15, Dieleman15, Kim17}.

ML allows us to express complex physical processes into simpler models. However, the speed and accuracy of ML come at the cost of a lack of understanding of the inner workings of the ML system and how this could map to physical processes. For some applications (such as photometric redshifts) this is not a major concern but for others, this \myquote{black box} approach can limit their applicability or even our ability to update the model when new physical understanding is gained. On the other hand, ML can be used not only as a black box but also to study the interplay between variables in a system, potentially leading to a better understanding of the physical processes involved \citep{Yip14, Hoyle15}.

In this paper, we compute different properties of dark matter haloes (assumed to host luminous galaxies) and apply ML techniques to classify haloes according to their cosmic environment and extract the most important properties that relate haloes to their environment. The ML contribution of this paper can be described as follows:  First, we provide a simple way of encoding the galaxy properties and history into feature vectors. Second, the technique described here provides a fast and computationally efficient galaxy-LSS classification that relies on simple descriptors such as local density, mass, formation time, merger history, etc. to accurately assign haloes to their cosmic environment. Third, we apply the Least Absolute Shrinkage and Selection Operator (LASSO) \citep{tibshirani1996regression} technique to identify the most significant galaxy properties that encode environmental effects, gaining a better understanding of what galaxy properties are relevant for environmental studies. Last but not the least, we observed a decent classification accuracy based on just information from the history of the haloes, which shows that environmental information is actually encoded in the history of the haloes in a measurable way.

The rest of the paper is organized as follows: Section 2 explains how the data was generated from N-body simulations. We describe our analysis and results in Section 3, followed by some closing remarks in Section 4. A detailed description of the techniques is included in the Appendix.

\section{Data}

\subsection{N-body simulations}

Our analysis is based on the MIP simulation \citep{Aragon16}. The MIP simulation consists of 256 realizations of a 32 $h^{-1}$Mpc box, each containing  $256^3$ particles, giving a mass per particle of $1.62\times10^{8}$M$_{\odot}$h$^{-1}$. 150 snapshots were evolved and stored at logarithmic intervals starting at $z=10$ until the present time using the publicly available N-body code Gadget \citep{Springel01}. We adopted a $\Lambda$CDM cosmology with parameters $\Omega_m = 0.3$, $\Omega_{\Lambda}=0.7$, $h=0.73$,
$\sigma_8 = 0.84$ and spectral index $n=0.93$, of the same order of values measured by the Planck mission \citep{Planck15}, the exact values are not relevant to the present work. The box size of the MIP is large enough to contain several cosmological voids and their surrounding walls and filaments. The largest cluster in the simulation is $\sim 10^{14}$ h$^{-1} $M$_{\odot}$. For the purposes of this paper, the MIP is sufficient in terms of size and number of available haloes.

From every snapshot in the simulation, we computed friends of friends (FoF) groups with a linking length of $b=0.2$ and physical properties such as mass, radius, shape, angular momentum etc.

\begin{figure}
\label{meg}

\subfigure{\includegraphics[clip, trim=7cm 0cm 7cm 0cm,width=0.16\textwidth,height=10cm,angle=0.0]{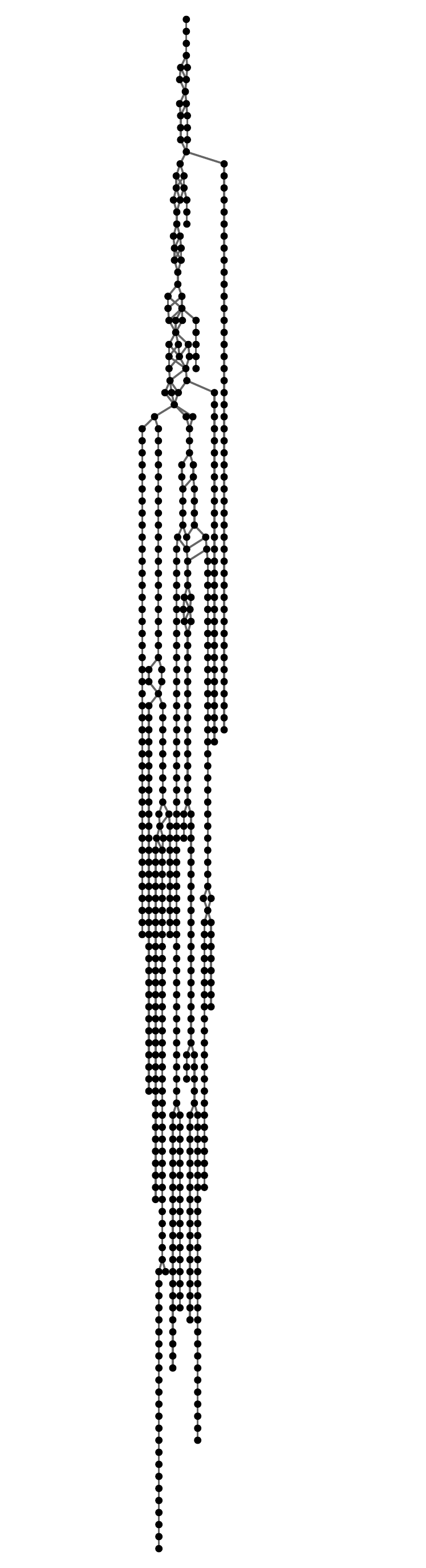}}
\subfigure{\includegraphics[clip, trim=7cm 0cm 7cm 0cm,width=0.16\textwidth,height=10cm,angle=0.0]{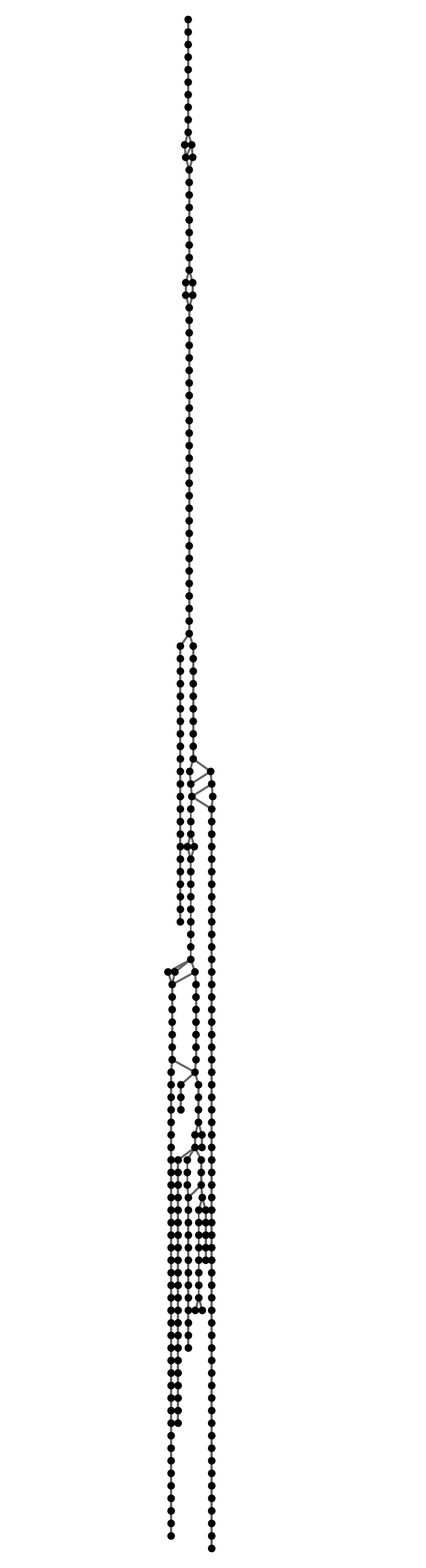}}
\subfigure{\includegraphics[clip, trim=0cm 2cm 0cm 0cm,width=0.13\textwidth,height=9.5cm,angle=0.0]{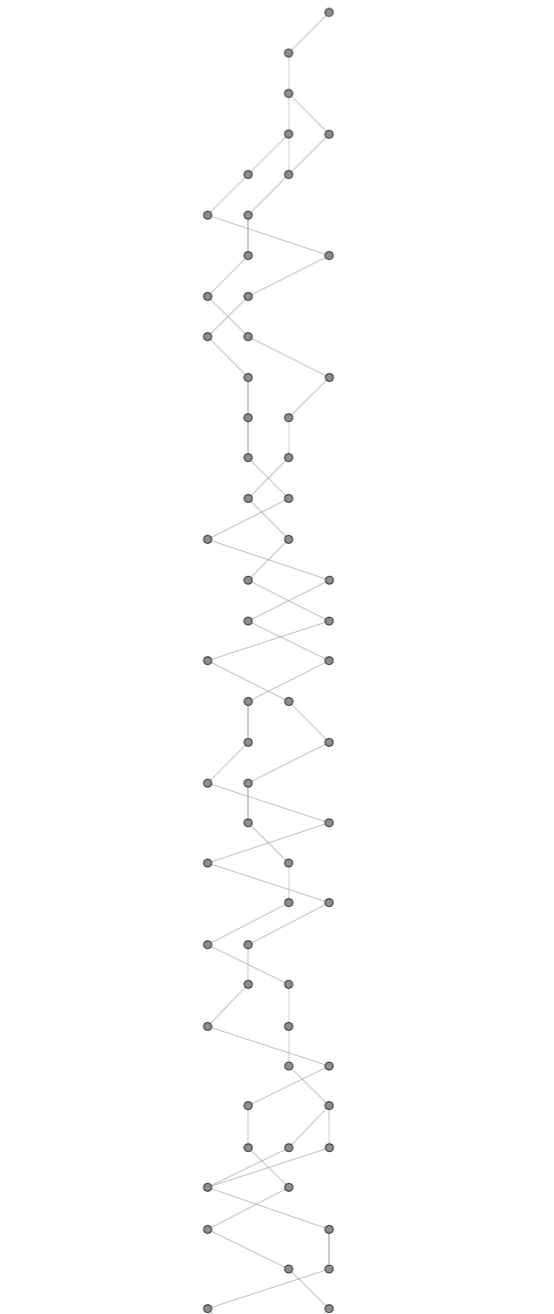}}
\caption{Merger Trees in an overdense environment (left panel), underdense environment (central panel) and a zoom inside a merger tree showing the individual nodes and their links. The dots correspond to haloes at different snapshots, here encoded in the vertical axis starting from the bottom (the last snapshot corresponds to the top of the tree). Links indicate mass transfer between haloes in different snapshots. The connectivity corresponds to a graph (since there are loops caused by mergers and other non-linear processes) but in general they have a tree-like structure.}
\label{merger tree}
\end{figure}

\subsection{Halo merger trees}

In order to follow the evolution of individual haloes we generated their merger trees as follows: For every FoF halo identified at snapshot $i$ (child halo), we map its particles into the previous snapshot $i-1$. The mapping operation is possible since every particle in the snapshot has a unique ID that is preserved throughout the simulation run. A halo in snapshot $i$ typically maps most of its particles into one halo in snapshot $i-1$ and several other halos with a small number of common particles. This operation is repeated for each of the progenitor halos in a recursive way until reaching the first snapshot in the simulation or until no more halos are found. The network representing the connections of a given halo with its progenitor haloes in previous snapshots is called a merger tree. Strictly speaking, the links connecting haloes across the simulation construct a graph (mainly arising from loops in the tree during halo merger events) and it is customary to prune the graph to produce a tree in which two haloes can not share the same progenitors. Fig. \ref{merger tree} shows a comparison of three merger trees within different LSS.

\subsection{LSS classification, the Spine method}

The LSS classification used as a reference and for training in our analysis was obtained with the Spine method \citep{Aragon10a} extended to a hierarchical formalism as described in \citet{Aragon10b}. The Spine method produces a full characterization of space into voids, walls filament, and clusters on a voxel basis. Haloes are assigned the Spine classification of the voxel where the halo's center is located. It is important to note that the LSS classification computed with the Spine method is based on the topology of the large-scale density field and is independent of the properties of haloes, thus providing an orthogonal method to the one proposed here.

From the $N$-body simulation we computed graph data of each of the 150 snapshots. The final dataset consists of 14975 haloes with merger trees. A visualization of merger trees is shown in Figure \ref{merger tree}. Along with the merger trees, the intrinsic properties of each halo such as mass and density with two different scales are given as well.

Local overdensities were computed inside a spherical tophat window with radius $r=1$h$^{-1}$Mpc and $r=2$h$^{-1}$Mpc centered at the position of each halo. The number of particles inside each tophat window was then divided by the mean number of particles inside the volume of the window. In the rest of the paper we refer to this quantity as \textit{density} to avoid confusion with the \textit{overdense} class.

%
\section{Analysis and Results}

In order to identify to which of a set of categories a new observation belongs, a crucial task in classification is finding an adequate representation that is able to describe the training data. This is usually done with the use of feature vectors $\bf{x} \in \mathbb{R}$ which are composed of features used to characterize the object. Representing objects by feature vectors can benefit from the mathematical wealth of operations available in vector space, which leads to algorithms with low computational complexity. In our case, the intrinsic properties of halos such as mass and density can easily be incorporated in the form of feature vectors. However, the difficulty remains in representing the merger trees.

Details of the feature extraction procedures are included in Section \ref{fex} and the classifiers are described in Section \ref{intro:classifier}.

\subsection{Merger Tree Feature Extraction}
\label{fex}

\begin{figure}
\includegraphics[clip, trim=5cm 0cm 5cm 0cm,width=0.20\textwidth,height=9cm,angle=0.0]{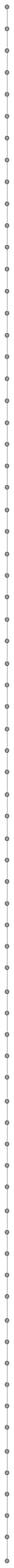}
\includegraphics[clip, trim=5cm 0cm 5cm 0cm,width=0.23\textwidth,height=9cm,angle=0.0]{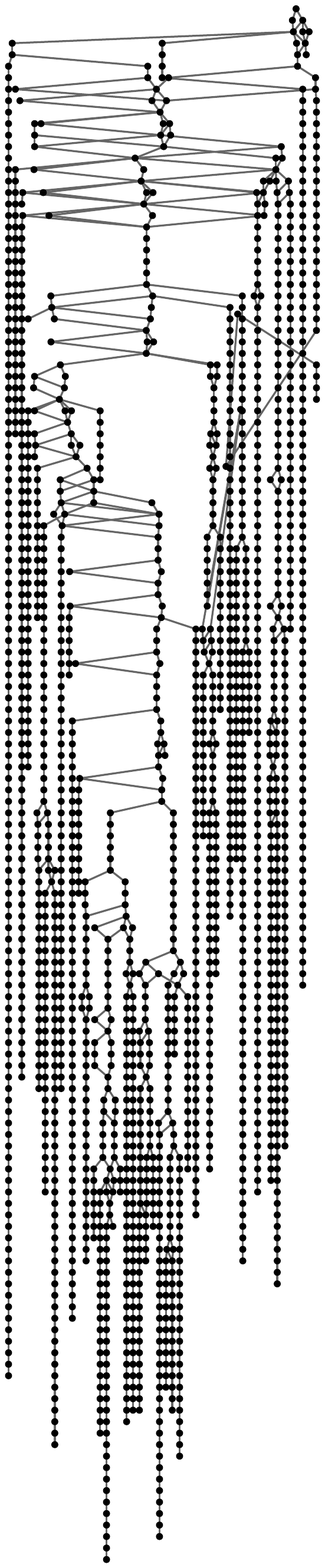}
\caption{{\bf Straight vs. Tree-like structure}. We call the left merger tree straight (straight $=1$) and the right one (straight $=0$) tree-like structure.}
\label{straight}
\end{figure}

Adapting existing feature extraction methods to halo merger trees data is nontrivial. We first observe that most of the haloes within the underdense class experience little changes (merge or split), leading their merger trees to look like straight lines (see Figure \ref{straight}). Therefore, we propose as a first indicator a feature representing the branches of the merger trees. We define the feature ``straight" as follows:

\begin{equation*}
\hbox{straight}=\left \{
\begin{array}{ll}
1, & \hbox{if the merger tree is branched}\\
0, & \hbox{otherwise}
\end{array}
\right. .
\end{equation*}

\noindent Note that the straight feature depends on the simulation's mass resolution. A higher mass resolution will result in more branches in the merger trees even for the \textit{straight} class. However, the straight feature is correlated with the cosmic environment.  Higher mass resolution will most likely map to very underdense regions in which case it will still have discriminating value.

%
\subsubsection{Algebraic Connectivity of Graphs}

The complexity of the merger history of haloes depends on their environment. Haloes in the underdense void-wall environment experience few encounters and mergers with other haloes compared to haloes in the denser filament-wall environment. This difference is reflected in the degree of complexity of their merger tree. We expect haloes in filament-wall environments to be complex or well-connected and haloes in wall-voids to have relatively simpler merger trees.

We first define some matrices that will be used throughout this paper. Let $G=(V, E)$ be a non-directed finite graph without loops and multiple edges. The {\bf{adjacency matrix}} of $G$ is defined as:
\begin{equation*}
A(i,j)=\left \{
\begin{array}{ll}
1, & \hbox{if} ~ (i,j) \in E \\
0, & \hbox{otherwise}
\end{array}
\right..
\end{equation*}
The {\bf{degree matrix}} $D$ for $G$ is a $n \times n$ diagonal matrix defined as:

\begin{equation*}
D(i,j)=\left \{
\begin{array}{ll}
\hbox{deg}(v_i), & i=j\\
0, & \hbox{otherwise}
\end{array}
\right.,
\end{equation*}
where $\hbox{deg}(v_i)$ is the number of edges attached to the vertex $v_i$. The {\bf{Laplacian matrix}} $L$ is defined as: $$L=D-A,$$ where $D$ is the degree matrix and $A$ is the adjacency matrix of graph $G$. Let $n \ge 2$ and $0=\lambda_1\le \lambda_2 \le \lambda_3 \le \cdots \le \lambda_n$ be the eigenvalues of the matrix $L$.  Applying the Perron-Frobenius theorem to $(n-1)I-L$, it follows that $\lambda_2$ is zero if and only if the graph $G$ is not connected. The second smallest eigenvalue $\lambda_2$ of the matrix $L(G)$ is called the algebraic connectivity of the graph $G$ in \citet{fiedler1973algebraic}. He also stated that the algebraic connectivity is a good parameter to measure, to a certain extent, how well a graph is connected. The algebraic connectivity is monotone: it does not decrease when edges are added to the graph.

The {\bf normalized Laplacian matrix} of $G$ is defined as:
$$\mathcal{L}=D^{-1/2}LD^{-1/2}$$
i.e.
\begin{equation*}
\mathcal{L}_{i,j}: =\left \{
\begin{array}{ll}
1, & \hbox{if} ~ i=j  ~ \hbox{and} ~ i \ne 0\\
\displaystyle -(d_i d_j)^{-1/2}, & \hbox{if} ~ (i,j) \in E\\
0, & \hbox{otherwise}
\end{array}
\right..
\end{equation*}

In \citet{chung1997spectral} the authors showed that the second smallest normalized Laplacian eigenvalue $\lambda_2^*$ of graph $G$ is 0 if and only $G$ is disconnected. In addition, Chung also established the relationships between $\lambda_2^*$ and the discrete Cheeger's constance and isoperimetric problems. Furthermore, $\lambda_2^*$ is also closely related to the aforementioned algebraic connectivity of $G$ (\cite{butler2008eigenvalues}). Thus, $\lambda_2^*$ is also known as a good parameter to measure how well a graph is connected. The authors in \cite{li2014six} classify trees into six classes $\mathscr{C}_1, \cdots, \mathscr{C}_6$ and prove that $\lambda_2(T_i)>\lambda_2(T_j)$ for $1 \le i < j \le 6$, where $T_i \in \mathscr{C}_i$ and $T_j \in \mathscr{C}_j$. More details of the second smallest eigenvalue of the normalized Laplacian matrix can be found in \cite{li2014six}.

We incorporate the second smallest eigenvalue of the normalized Laplacian matrix as the feature to measure the graph connectivity of the merger trees.

\begin{figure}
\centering
\includegraphics[width=0.49\textwidth,angle=0.0]{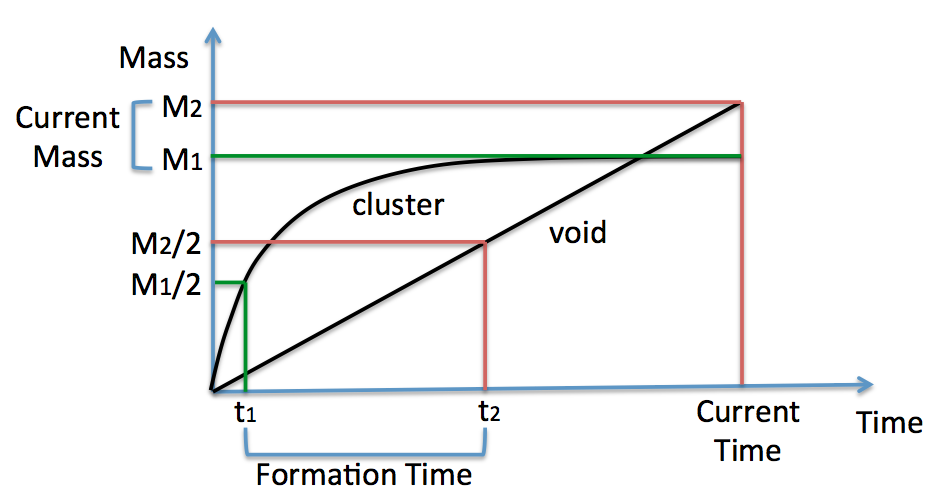}
\caption{{\bf Definition of Formation Time} Suppose we have two haloes $G_1$, $G_2$ and their mass $M_1$ and $M_2$ at current time, then the formation time for $G_1$ and $G_2$ is $t_1$ and $t_2$ which corresponds to half of their current mass ($M_1/2$ and $M_2/2$), respectively. Time is represented by snapshot number.}
\label{formtime}
\end{figure}

\subsubsection{Formation Time}
We define halo formation time as the snapshot number when the halo reaches half of its current mass. See Figure \ref{formtime} for the definition of the formation time. Based on the definition of formation time, we first find the most massive progenitor at each snapshot, then locate the one that has mass closest to half of the halo's current mass and mark the snapshot number as the formation time of the halo.

\subsubsection{Encoding the Tree-like structure}
The idea of developing the tree encoding technique was motivated by the work in \citet{4591399}, where the depth-first string encoding and the Prufer encoding were adopted to represent trees in appropriate forms to facilitate similarity searches and further classification. Since the number of snapshots is the same for all haloes, we consider a simple way to encode the tree-like structure by an $n \times 1$ vector, where $n$ is the number of snapshots. Start tracing the merger trees from present time to the past, each element represents the number of nodes the tree-like structure has at each snapshot. Not all the haloes have its progenitor in all $n$ snapshots, we simply fill in 0 when there is none. Since not all the haloes have the same formation time, we need to shift the merger tree with respect to their formation time. Shifting procedures are depicted in Appendix \ref{app:shift}. Another feature named depth is defined by the number of snapshots where the halo appears. In addition, we extract features ``ratio" and ``diff" from the merger trees where $$\mbox{ratio}=\frac{\mbox {total number of nodes}}{\mbox {total number of edges}}, ~\mbox{and}$$ $$\mbox{diff}={\mbox {total number of edges}}-{\mbox {total number of nodes}}.$$ Note that the range of the variable ratio will be (0,1) and larger values of the ratio are indicative of more complicated trees. A summary of the features is presented in Table \ref{features}.

\begin{table}
\centering
\caption{Summary of Features}
\label{features}
\begin{tabular}{lll}
\hline
Feature        & Description                                   & Range  \\
\hline
density        & tophat density                              & 0-1000 \\
mass           & halo mass                                     &    $10^{10}$-$10^{14}h^{-1}$M$_{\odot}$     \\
straight       & indicator of tree branching         & 0 or 1 \\
formtime & time to reach half of final mass & 1-151  \\
fidval         & Fiedler value        &   [0,2]     \\
depth          & length of the tree                            & 1-151  \\
shape         & $\displaystyle\frac{\text{smallest semi-axis}}{\text{largest semi-axis}}$         & [0,1]  \\
spin         & magnitude of the spin parameter                            & $(0,\infty)$ \\
ratio          & ratio node $\#$ \& edge $\#$          & (0,1)  \\
diff          & difference of edge $\#$ \& node $\#$          & (0,1)  \\
L1-L151        & node $\#$ at each level(shifted)   & 0-1000 \\
\hline
\end{tabular}
\end{table}

%
\subsection{Introduction of Classifiers}\label{intro:classifier}

In this section, we review the basic idea of support vector machine (SVM) and SVM with Distance matrix LU decomposition (LU-SVM).

%
\subsubsection{Support Vector Machine}

The support vector machine classifier is a binary classier algorithm developed to solve pattern recognition problems (\cite{boser1992training}; \cite{MR1641250}; \cite{cristianini2000introduction}). With this classifier, one maps the data into a higher dimensional input space and constructs an optimal separating hyperplane in that space that can maximize the gap between data points on the boundaries, allowing us to separate data points into classes. Fundamentally, finding the optimal separating hyperplane is equivalent to solving a quadratic programming problem (see Appendix \ref{app:svm} for details).

\subsubsection{Distance Matrix LU Decomposition}
We have now achieved a feature vector for each halo which contains the mass, density, normalized Fiedler value, straight, formation time, depth, ratio, diff and node number at each level of the merger tree. Denote a halo profile\footnote{Note that \myquote{profile} in this context is different than the traditional term used to describe the light profile} by $G_i=(g_1, g_2, \cdots, g_n)$, where $g_i$'s represent the features we extracted from the merger tree as well as intrinsic properties of the haloes and $n$ is the number of features, for $1 \le j \le n$. Let $G=(G_1, G_2, \cdots, G_N)^T$ be a set of all haloes where $N$ is the total number of haloes. Thus feature vectors of each halo under the same condition are denoted by a row vector of the matrix $G$ and all the halo profiles can be considered as the points mapped to a high-dimensional space and all the information is included in matrix $G$.

As the distance matrix is constructed with elements describing the space between points, we found that through the construction of the distance matrix, the relationship between various haloes can be reflected \citep{5620686}. Therefore, we defined the distance matrix $\mathcal{D}$ as follows:

\begin{equation}
\mathcal{D}_{i,j}: =\left \{
\begin{array}{ll}
||G_i-G_j||, &  ~ i \ne j \\
0, & ~ i=j
\end{array}
\right.,
\label{dis}
\end{equation}

where $0 \le i,j \le N$ and $G_i, G_j$ are the feature vectors of halo $i$ and $j$, respectively. By doing so, we are transforming the feature vectors into the space between the points, which represents the difference between various haloes. Then we factor the distance matrix as the product of a lower triangular matrix $L$ and an upper triangular matrix $U$ ($LU$ decomposition), where $L$ captures difference coefficient information between halo profiles and $U$ records different feature information of each halo profile after removing the redundant information in all the halo profiles. Now that all the different information between samples are stored in $L$ and $U$, we further integrate $L$ and $U$ into a matrix $H$ by $H=L+U-I_N$ where $I_N$ is the $N$-dimensional identity matrix. Then we take each row vector of $H$ as our updated feature vectors for each halo profile. In order to make the updated feature vectors comparable, we perform the following normalization to $l=(l_1,\cdots,l_j,\cdots,l_N)^T$ so that all the values are mapped to $[-1,1]$:
\begin{equation}
l^*_j=2\frac{l_j-\displaystyle \min_{1\le i \le N} l_i}{\displaystyle \max_{1\le i \le N} l_i-\min_{1\le i \le N} l_i}-1, j=1, \cdots, N.
\label{norm}
\end{equation}

To summarize, the steps of the LU-SVM algorithm are as follows:
\begin{description}
\item[1:] Input the feature vectors for all the haloes as row vectors into matrix $G$.
\item[2:] Use equation \ref{dis} to compute the distance matrix $\mathcal{D}$.
\item[3:] The distance matrix matrix $\mathcal{D}$ is then decomposed by $LU$ decomposition to obtain $L$ and $U$.
\item[4:] Integrate $L$ and $U$ into a matrix $H$ by $$H=L+U-I,$$ where $I$ is the identity matrix. Normalize H by equation (\ref{norm}).
\item[5:] Use row vectors of normalized $H$ as the classifiers' input and carry out two-fold, five-fold and leave-one-out cross validation.
\end{description}

The simulation results are shown in Table \ref{SVM}. The classification accuracy improved significantly (more than $20\%$ on average) with the application of LU decomposition.

\begin{table}
\centering
\caption{The classification results with leave-one-out cross validation, five-fold cross validation and two-fold cross validation are reported (standard errors are given in parenthesis). Each of the accuracy values reported are based on 100 random sampling of size 1400 with equal probability from each class.}
\label{SVM}
\begin{tabular}{|l|c|c|c|l}
\cline{1-4}
SVM                  & Leave-one-out & Five-fold    & Two-fold     &  \\ \cline{1-4}
All features         & 68.53(1.17)  & 67.89(1.24) & 66.46(1.34) &  \\ \cline{1-4}
Reduced features     & 75.23(1.11)  & 74.97(1.14) & 74.25(1.31) &  \\ \cline{1-4}
Merger tree features & 61.01(1.11)  & 60.83(1.14) & 60.64(1.18) &  \\ \cline{1-4}
\end{tabular}

\begin{tabular}{|l|c|c|c|l}
\cline{1-4}
LU-SVM               & Leave-one-out & Five-fold    & Two-fold     &  \\ \cline{1-4}
All features         & 93.32(0.74)  & 92.82(0.75) & 91.65(0.91) &  \\ \cline{1-4}
Reduced features     & 93.29(0.62)  & 92.89(0.67) & 92.02(0.86) &  \\ \cline{1-4}
Merger tree features & 79.94(1.61)  & 79.12(1.64) & 77.39(1.96) &  \\ \cline{1-4}
\end{tabular}
\end{table}

\begin{figure}
\centering
\includegraphics[width=0.52\textwidth,height=4cm,angle=0.0]{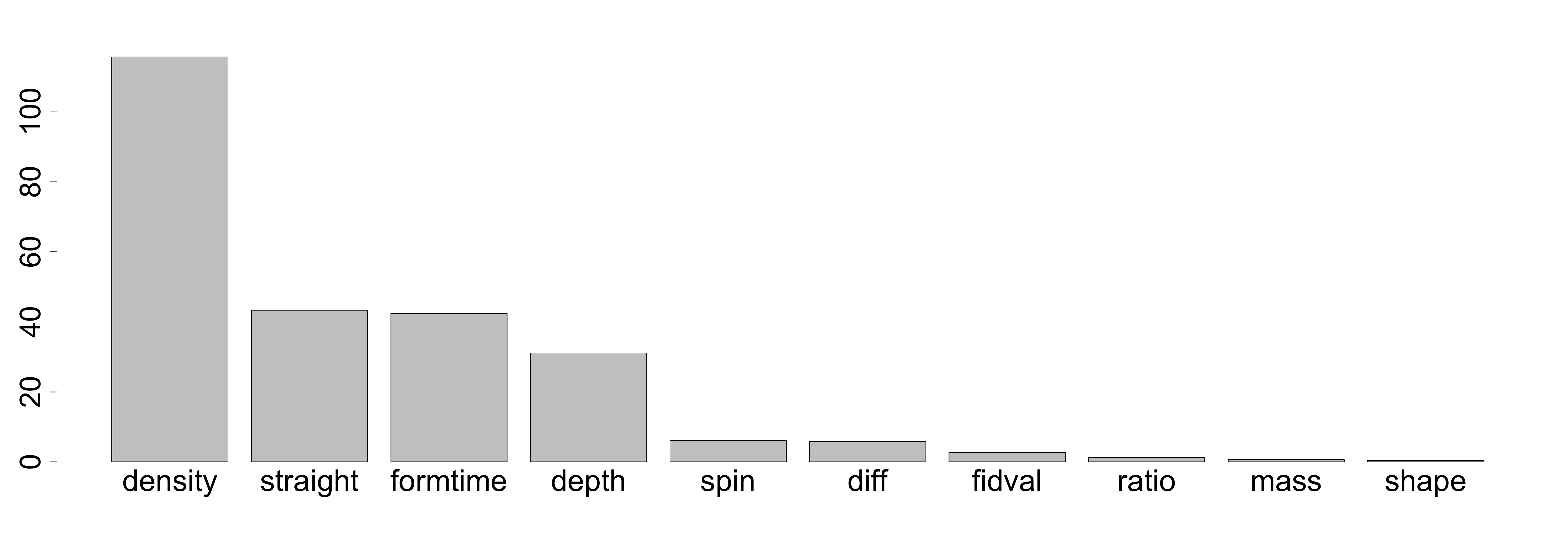}
\caption{Feature scores for the ten most significant features. The scale on the vertical axis is arbitrary. Density is clearly the most significant feature followed by formation time, straight and depth with almost equal feature scores. The remaining features have less than one-third of the maximum feature score.}
\label{featurescore}
\end{figure}

\subsection{Feature Selection and Feature Importance}\label{sec:feature importance}

By far, we have extracted the following features: density and mass (from halo intrinsic properties); straight, formation time, Fiedler value, depth, ratio, difference and node number at each level (from the merger tree). From the set of features used to describe a halo and its history, we should expect that only a small number of features would contain relevant information on the cosmic environments of haloes. From Fig. \ref{density} it is clear that local density alone can provide a first-order LSS classification. However, the overlapping tails between the void/wall and filament/cluster mass density distributions (see Fig. \ref{density}) mean that density alone can not fully separate between the halo populations.  The straight measure also seems intuitive to differentiate between haloes in dynamically young environments, in which case their merger tree would be very simple, and haloes in dynamically evolved environments where we expect to see haloes with complex merger trees.

In order to provide a quantitative measure of the importance of each feature in the LSS classification, we performed a feature selection with the Least Absolute Shrinkage and Selection Operator (LASSO) method (\cite{tibshirani1996regression}). The Lasso is a shrinkage and selection method for linear regression. It is often adopted for variable selection and regularization in order to increase the prediction accuracy and interpretability. To be specific, several random subsets are sampled from the dataset and logistic regression models are fitted for each random subset. A score is then assigned to each feature based on the tendency of LASSO in including that feature in the models. Figure \ref{featurescore} shows that out of the 161 features used in the full analysis only 4 have relatively high feature score: density, straight, formation, and depth. As expected density is the most important feature and the effect of the remaining three features somehow encode aspects of the cosmic environment to which density is insensitive. If we use only straight, formation time and depth to classify haloes then we obtain an accuracy of 61.01\% with a standard deviation of 1.11\%, which is slightly above the random expectation but high enough to improve the classification when used in addition density from 61.01\% to 75.23\%. In fact, if we apply LU-SVM, the accuracy went up to 93.29\%. We can see that the performance with the four most significant features is almost as good as using the whole feature set. This is reasonable considering the gist of LU-SVM is to capture the differences between haloes (as the fundamental matrix is built with the distance between halo feature vectors). Furthermore, this indicates that the four most significant features could help to reveal the hidden processes of how the halo formation is associated with the large-scale structure.

\section{Closing Remarks}

We applied a set of machine learning techniques in order to gain a better understanding of the processes behind halo formation and to provide with a computationally fast algorithm to classify haloes according to their cosmic environment. It takes about three minutes in a regular workstation to classify the haloes with LU-SVM. The techniques presented here enables us to embed merger trees into feature vectors and infer large-scale structure based on them. On top of that, we adopted the LASSO method for logistic regression on the full set of features. With a set of tuning parameters $\lambda$ being supplied, we were able to reduce the features and achieve similar prediction accuracy. We selected four most significant features that are associated with the large-scale structures and found that them alone can already classify haloes into void/wall or filament/cluster with the accuracy of 93\%.

The fact that halo properties themselves, apart from local density, contain an imprint of their cosmic environment is remarkable, even if the measured signal is small. Standard models of halo formation assume that all the information needed to reproduce the properties of haloes is contained in their mass accretion history, computed via their merger tree, while the cosmic environment is assumed to be implicit in the merger tree information. The studies presented here show in a quantitative way that indeed there is environmental information encoded in the merger tree of haloes and that this information can be exploited to derive the position of haloes in the Cosmic Web.

We found the mass of a halo to be a poor indicator of its cosmic environment. This could be the result of the strong overlap between the mass function of the two classes explored here. For the range of masses, we explored the mass function is very similar and impossible to separate in feature space as in the case of the density distributions where we have two clearly separated peaks. The mass function in underdense and overdense environments only differs in the slope of the mass range studied here (roughly around and below M$\ast$). For very massive haloes mass should play an important role, the simulation used in this work contain only a handful of haloes more massive than  $>10^{14}$ \msun and therefore their contribution to the total classification is small.

While the results presented here are dependent on the resolution of the N-body simulation, the general trends should be the same once appropriate constraints are applied to higher-resolution simulations. This is particularly important for our definition of \myquote{straight} vs. \myquote{complex} since a simulation with a higher mass resolution (i.e. more particles used to define the same halo) will most likely transform the \myquote{straight}  trees into \myquote{complex}. However, the branches in such tree will consist of low-mass objects that could be easily discarded. Our present work is robust in the sense that the mass resolution used to define \myquote{haloes} is sufficient to resolve halos with a mass corresponding to the smallest galaxies observable in galaxy surveys.

The method presented here can be easily applied to existing N-body simulations with minimal modifications and computational cost since N-body simulations routinely compute halo/subhalo catalog and their merger trees. Computing all the features needed for our method is straightforward, and as discussed in Sec. \ref{sec:feature importance}, we may only need to compute a handful of features with the largest contribution to the LSS classification. This makes our method attractive for large-scale cosmological simulations with number of particles of the order of trillions and large computational boxes which make grid-based LSS analysis methods computationally expensive. The use of merger trees and simple halo properties allows LSS classifications to be computed for each timestep/snapshot of the simulation, providing a fine-scale classification at minimal computation.

As a point of comparison, the Spine method used to classify the simulation used in this work requires the computation of the density field inside a regular grid. Given the relatively small box size (32 \mpc) a grid of  $256^3$ voxels is more than sufficient to sample the LSS elements and assign haloes to them. However, for a box of 1 Gpc, we would require a grid size of $2000^3-4000^3$ voxels (for a resolution of $0.5-0.25$ Mpc/voxel) and this may still not be enough to avoid assigning haloes to a wrong LSS element if a halo is sitting near the boundary between different LSS elements.

The use of ML techniques makes our method intrinsically resolution-independent as it depends only on the training data. The method can be applied to simulations of any size as long as they have similar mass resolution which means we can train on a small simulation and apply it to a much larger one without modification. The need for similar mass resolution between the training data and the target data could be seen as a potential limitation. However, given that we can train on a modest simulation in practice, this is irrelevant.

This is the first of a series of papers exploring the features that determine the properties of haloes and their relative importance. Machine learning techniques provide a quantitative way to assess and to understand the interplay between different halo properties in an area dominated by qualitative and semi-empirical studies.

\vfill

\bibliography{refs}
\bibliographystyle{mn2e}

\appendix

\section{Support Vector Machine} \label{app:svm}
Given training data set $\{{\bf{x_k}},y_k\} \in \mathbb{R}^n \times \{-1,1\}$, where $\bf{x}_k$ are feature vectors and $y_k$ the class labels. For linearly non-separable case, the feature vector $\bf{x}$ is mapped into a high dimensional feature space by function $\phi$. This is often referred as ``the kernel trick". Then the SVM discriminant function has the form:
$$f({\bf x})=\hbox{sign} [{\bf w}^T \phi({\bf x})+b],$$
where $\bf w$ is the parameter vector, b is the offset scalar and $\phi(\cdot)$ is a nonlinear function that maps the input space into a higher dimensional space (can be infinite dimensional).

For separable data, the assumptions are:
\begin{equation*}
\left \{
\begin{array}{ll}
{\bf{w}}^T{\bf \phi(x_k)}+b \ge 1, &  ~\hbox{if} ~~ y_k=1  \\
{\bf{w}}^T{\bf \phi(x_k)}+b \le -1, &  ~\hbox{if} ~~ y_k=-1
\end{array}
\right.,
\end{equation*}
which is equivalent to
$$y_k[{\bf w}^T{\bf \phi(x_k)}+b] \ge 1, ~k=1,\cdots,N.$$
To handle non-separable datasets, the constraints were relaxed by making the inequalities easier to satisfy. Slack variables $\xi_i \ge 0$ were included:
$$y_k[{\bf w}^T{\bf \phi(x_k)}+b] \ge 1-\xi_k, ~k=1,\cdots,N.$$
All constraints can be satisfied if $\xi_k$ is large enough trivially. To prevent this, the sum of $\xi_k$ was added as a penalty and the optimization problem becomes:
\begin{equation*}
\begin{aligned}
\text{minimize:} ~~&\frac{1}{2} \Vert{\bf w}\Vert^2 + C \sum_{i=1}^n \xi_k, \\
\text{subject to:} ~~&y_k[{\bf w}^T{\bf \phi(x_k)}+b] \ge 1-\xi_k ~\text{and}~\xi_k \ge 0\\
&\text{for $k=1,\cdots,N.$}
\end{aligned}
\end{equation*}
where $\Vert \cdot \Vert$ is $l_2$ norm of a vector and $C$ is a weight parameter that needs to be chosen with cautious. Then constructs the Lagrangian
\begin{equation*}
\begin{aligned}
&\mathcal{L}({\bf w}, b, \xi; \alpha, \nu)=\frac{1}{2} ||{\bf w}||^2 + C \sum_{i=1}^n \xi_k-\sum_{k=1}^N \alpha_k\{y_k[{\bf w}^T{\bf \phi(x_k)}+b]\\
&-1+\xi_k\}-\sum_{k=1}^N\nu_k \xi_k
\end{aligned}
\end{equation*}
by introducing Lagrange multipliers $\alpha_k \ge 0, \nu_k \ge 0 (k=1,\cdots,N)$. The solution is given by the saddle point of the Lagrangian by computing
$$\max_{\alpha_k,\nu_k} \min_{{\bf w}, b, \xi_k}\mathcal{L} ({\bf w}, b, \xi_k; \alpha_k, \nu_k).$$
One obtains
\begin{equation*}
\left \{
\begin{array}{ll|||}
\displaystyle\frac{\partial \mathcal{L}}{\partial {\bf w}}=0 \rightarrow {\bf w}=\sum_{k=1}^N \alpha_k y_k \phi({\bf x}_k)\\
\displaystyle\frac{\partial \mathcal{L}}{\partial b}=0 \rightarrow \sum_{k=1}^N \alpha_k y_k=0\\
\displaystyle\frac{\partial \mathcal{L}}{\partial \xi}=0 \rightarrow
0 \le \alpha_k \le c, k=1,\cdots,N.
\end{array}
\right.
\end{equation*}
By solving the dual problem and introducing Lagrange multipliers, we arrive at
\begin{equation*}
\begin{aligned}
\text{maximize:} ~~& -\frac{1}{2} \sum_{i,j=1}^N \alpha_i \alpha_j y_i y_j \phi({\bf x_i})^T \phi({\bf x_j}) + \sum_{i=1}^N \alpha_i, \\
\text{subject to:} ~~&0 \le \alpha_i \le C ~\text{and}~\sum_{i=1}^N \alpha_i y_i=0\\
&\text{for $i=1,\cdots,N.$}
\end{aligned}
\end{equation*}

Then we obtain the classifier:
$$f({\bf{x}})=\hbox{sign}\left[\sum_{k=1}^N \alpha_k y_k \Phi({\bf{x,x_k}})+b\right],$$
where $\alpha_k$'s are positive real constants and $b$ is a real constant. Under Mercer condition,
$$\Phi({\bf{x_i,x_j}})=\phi(x_i)^T\phi(x_j).$$ The kernel function applied here is the RBF SVM:
$$\Phi({\bf{x,x_k}})=\exp\{-||{\bf{x-x_k}}||_2^2/\sigma^2\}.$$

\section{Shifted merger tree with respect to formation time}\label{app:shift}

\begin{figure*}
\label{shiftform}
\begin{center}
\includegraphics[height=6cm,width=7cm]{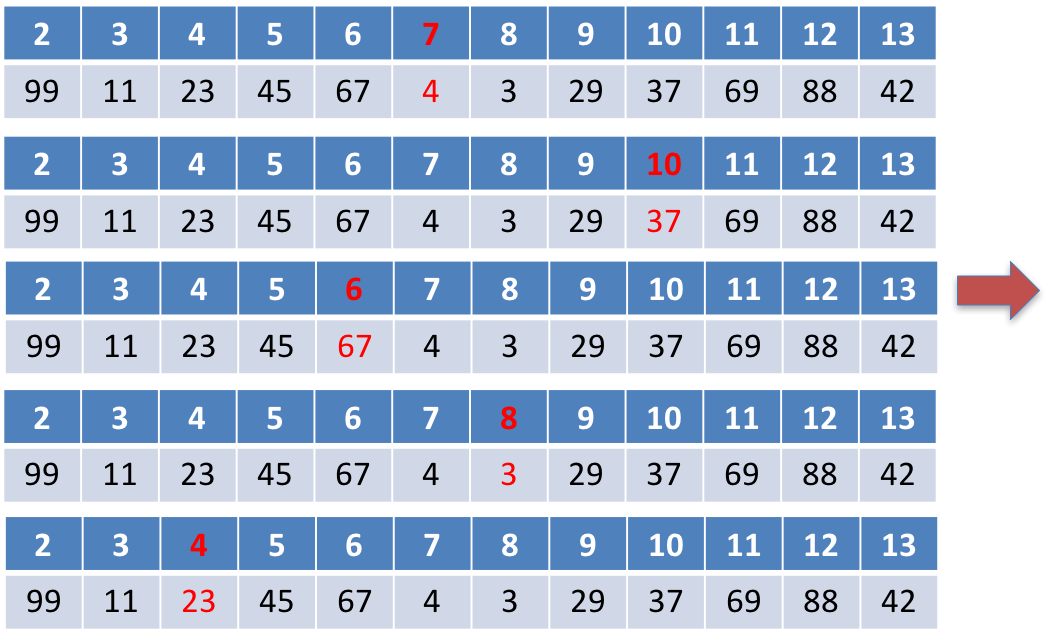}
\includegraphics[height=6cm,width=9cm]{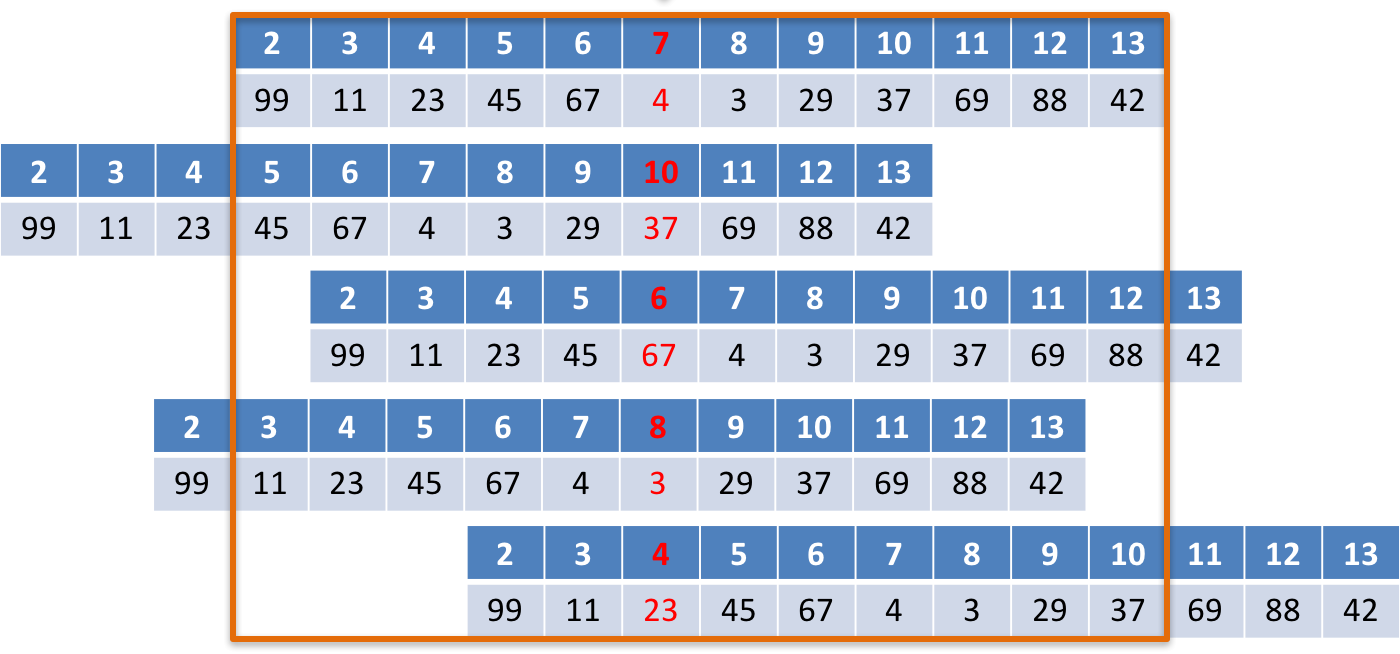}
\includegraphics[height=6cm,width=7cm]{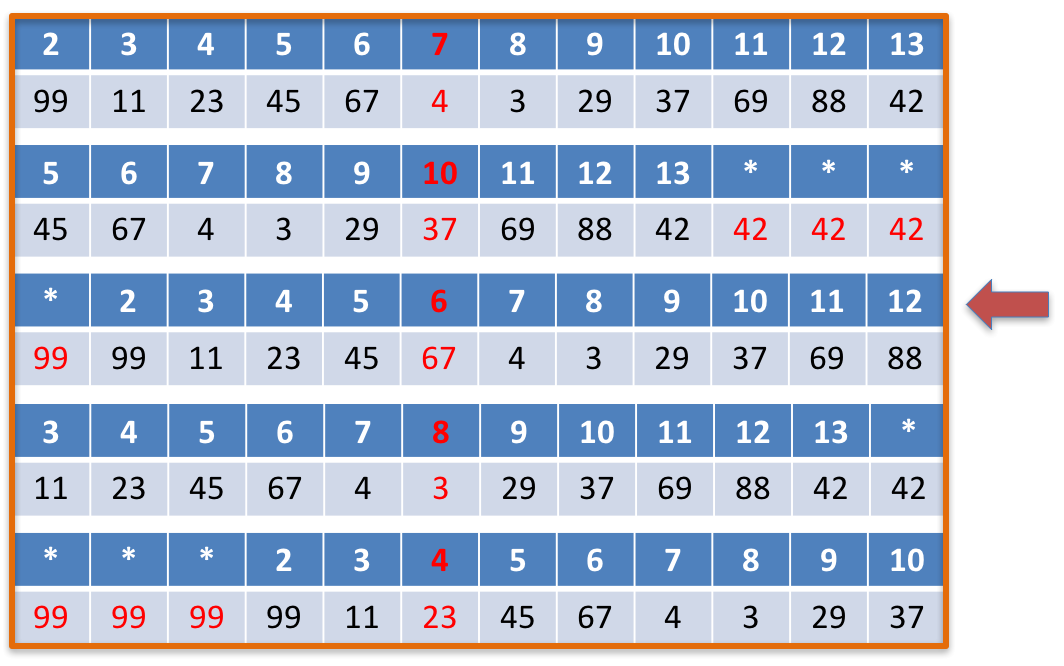}
\includegraphics[height=6cm,width=9cm]{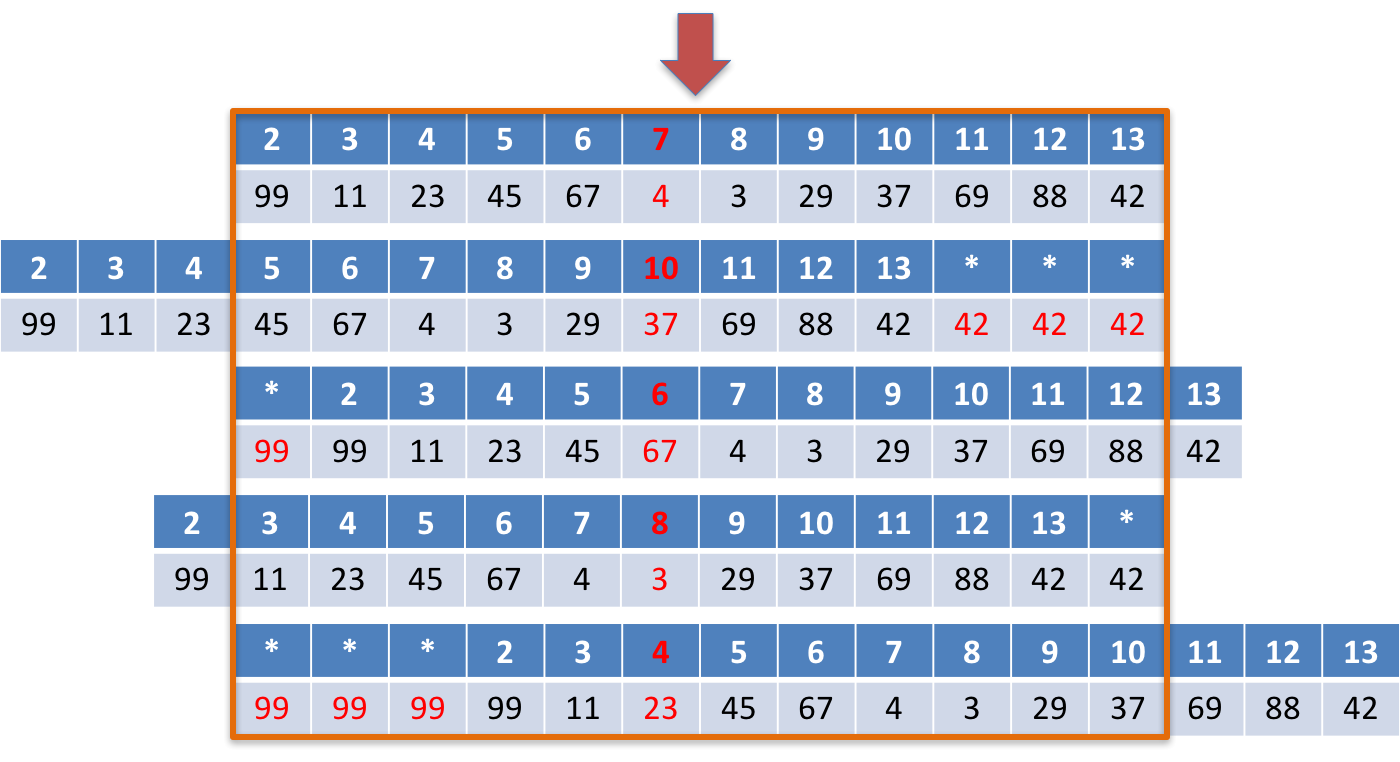}
\includegraphics[height=6cm,width=7cm]{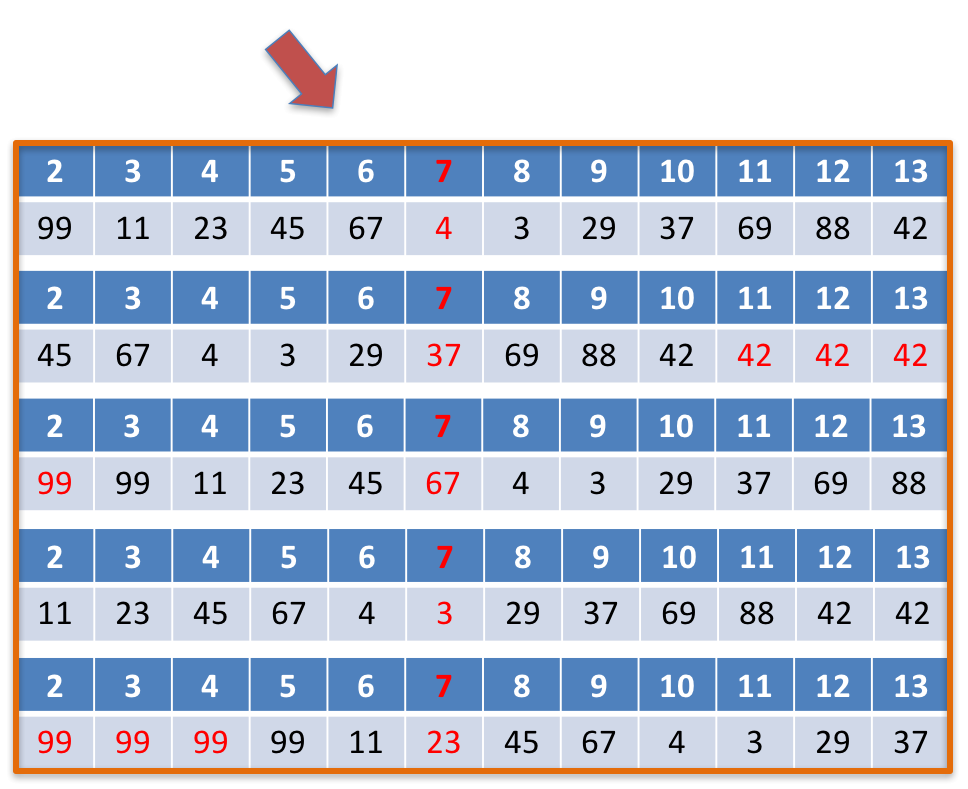}
\caption{{\bf{Demo of shifting merger tree with respect to the formation time}}. In all subplots, we are showing five haloes with different formation time (marked in red). The dark blue cells represent the snapshot number and the light blue cells contain the corresponding node number.}
\label{shifted}
\end{center}
\end{figure*}

Note that halo formation time is defined as the snapshot number when the halo reaches half of its current mass (see Figure \ref{formtime} for details). Consider formation time as a time marker for all the haloes, it is logical to shift the merger tree with respect to formation time so that the corresponding haloes are comparable to each other. The steps are as follows:

\begin{enumerate}
\item Locate the formation time and shift the merger tree with respect to the formation time. (See the first two subplots in figure \ref{shifted}).
\item Fill in the blank cells as a continuation of the adjacent number. 
\end{enumerate}

A demo is included in Figure C1.

\section{The LASSO technique}

The Lasso, introduced by Robert Tibshirani (\cite{tibshirani1996regression}), is a shrinkage and selection method for linear regression. It minimizes the residual sum of the squared subject to the sum of the absolute values of the coefficients being less than a constant. Because of the constraint, the lasso method is often adopted for variable selection and regularization in order to increase the prediction accuracy and interpretability.

Consider a sample with $n$ observations, each of which consists of $p$ covariates and a single outcome. Let $y_i$ be the outcome and $x_i :=(x_1,x_2,\cdots,x_p)^T$ be the covariate vector for the $i^{\text{th}}$ observation.

Letting $\hat{\beta}=(\hat{\beta_1},\cdots,\hat{\beta_p})$, the lasso estimate $(\hat{\alpha},\hat{\beta})$ is defined by
\begin{equation*}
\begin{aligned}
    &(\hat{\alpha},\hat{\beta})=\arg\min\left\{\sum_{i=1}^n\left(y_i-\alpha-\sum_j\beta_j x_{ij}\right)^2\right\} \\
    & \text{subject to} \sum_j |\beta_j|\le t.
\end{aligned}
\end{equation*}

Here $t \ge 0$ is a tuning parameter. Let $\hat{\beta}_j^o$ be the full least squares estimates. If $t>\sum_{j=1}^p|\hat{\beta}_j^o|$, then the lasso estimates will be the same as the ordinary least squares estimates. Values of $t<\sum_{j=1}^p|\hat{\beta}_j^o|$ will cause shrinkage of the solutions towards 0, thus some coefficients may be exactly equal to 0. Then the problem is equivalent to

\begin{equation*}
\begin{aligned}
    (\hat{\alpha},\hat{\beta})=\arg\min\left\{\sum_{i=1}^n\left(y_i-\alpha-\sum_j\beta_j x_{ij}\right)^2+\lambda\sum_j |\beta_j|\right\}
\end{aligned}
\end{equation*}

A penalty term $\lambda\sum_j |\beta_j|$ is added to the loss function. Each non-zero coefficient adds to the penalty, which forces weak features to have zero as coefficients. It has been shown in the same article that $\lambda$ depends on the LASSO parameter $t$, i.e. larger $\lambda$ yields smaller numbers of selected features.

\vfill

\end{document}